\documentclass[useAMS]{mn2e}
\usepackage{graphics,graphicx,array,amssymb}

\newcommand{\arcdeg}{\ensuremath{^{\circ}}}

\title[Trend Filtering Algorithm]
{A Trend Filtering Algorithm for wide field variability surveys}

\author[G. Kov\'acs, G. Bakos and R. W. Noyes]
{G\'eza Kov\'acs$^{1}$\thanks{E-mail: kovacs@konkoly.hu (GK); 
gbakos@cfa.harvard.edu (GB); noyes@cfa.harvard.edu (RWN)}, 
G\'asp\'ar Bakos$^{2,1\star}$\thanks{Predoctoral Fellow}
and Robert W. Noyes$^{2\star}$
\\
$^1$Konkoly Observatory, P.O. Box 67, H-1525 Budapest, Hungary\\
$^2$Harvard-Smithsonian Center for Astrophysics, 60 Garden Street, 
    Cambridge, MA 02138
}
\begin{document}

\date{Received day month 2004 / Accepted day month 2004}

\pagerange{\pageref{firstpage}--\pageref{lastpage}} \pubyear{200?}

\maketitle

\label{firstpage}

\begin{abstract}
We show that various systematics related to certain instrumental effects 
and data reduction anomalies in wide field variability surveys can be 
efficiently corrected by a Trend Filtering Algorithm (TFA) applied to the 
photometric time series produced by standard data pipelines. Statistical 
tests, performed on the database of the HATNet project, show that by the 
application of this filtering method the cumulative detection probability 
of periodic transits increases by up to 0.4 for variables brighter than 
11~mag with a trend of increasing efficiency toward brighter magnitudes. 
We also show that TFA can be used for the reconstruction of periodic 
signals by iteratively filtering out systematic distortions. 
\end{abstract}



\section{Introduction}
In the resolute hunt for transiting hot Jupiters, much effort is 
focused on very small instruments that are capable of observing 
large area of the sky and gathering data with a photometric precision 
better than 0.01~mag. There are many projects targeting this goal 
(see Horne 2003), using instruments with typical apertures of order 
10~cm; it has even been argued that the small telescope projects 
with $\sim 2$~inch diameter optics are the best for this purpose 
(Pepper, Gould \& Depoy 2003). 

The attained photometric precision plays a central role in transit 
signal detection. However, the large field of view and the 
semi-professional CCD detectors often employed by these low-budget 
projects drastically amplify some of the standard problems of CCD 
photometry. These include narrow PSF, crowded field, differential 
extinction and refraction (see Bakos et al.~2004, hereafter B04, 
for a more comprehensive summary). Most of the problems become more 
severe toward fainter magnitudes, where the objects are more affected 
by time-dependent merging with brighter stars. Most of these, and 
other hidden effects not only increase the noise of the light curves, 
but leave their fingerprint in the data as systematic variations,
because they vary in a non-smooth fashion over the field, and standard 
position-dependent smooth transformations applied at various stages 
of the data reduction do not eliminate them perfectly. 

The efficiency of the detection of periodic signals drops dramatically 
when the light curves become dominated by noise {\em and} systematic 
variations. For example, the BLS method (Kov\'acs, Zucker \& Mazeh 2002, 
hereafter K02), an application used for searching for transits, is 
efficient on light curves exhibiting a box-shaped dip, but fails to 
recover transits superposed on strongly variable light curves. These 
systematics are also harmful in discovering interesting low-amplitude 
phenomena, such as $\delta$~Scuti or $\gamma$~Dor stars. 

Many of the systematic variations in a given light curve are shared 
by light curves of other stars in the same data set, due to common 
effects like colour-dependent extinction, or blending of two or more 
star images (which could produce similar light curve variations in 
all of them), etc. A solution to remove or reduce these common 
systematic variations can therefore be devised using the already 
reduced data. For each target star, one must identify objects in the 
field that suffer from the same kind of systematics as the target, 
and apply some kind of optimum filtering of the target light curve 
based on the light curves of a set of comparison/template stars.

The algorithm to be presented in this paper can be applied to two 
types of problems. The first application is to remove trends from 
trend- and noise-dominated time series, thereby increasing the 
probability of detecting a weak signal (periodic or non-periodic). 
Secondly, if there {\em is} a periodic signal present, the TFA 
algorithm can be used in a slightly different way to reconstruct 
also the shape of that signal. 

The structure of the paper is the following. Section 2 gives a more 
detailed description of the systematics and their possible causes. 
Section 3 describes the mathematical formulation of the TFA 
algorithm, and Section 4 presents a number of tests of its ability 
to remove spurious trends from the data.  Section 5 explores the 
increase in detectability of periodic transit signals (using the 
BLS algorithm) from data after processing with the TFA. Section 6 
then investigates the capability of reconstructing the detailed 
shape and amplitude of such periodic signals, using a variant of 
the basic algorithm. The paper closes with a brief highlighting 
of the most important results. Throughout this paper we rely on 
the database of the HATNet project (see B04).


\section{Light curve systematics and their possible sources}
The systematic variations might be intrinsic to the data, or they 
could be due to uncorrected instrumental effects or changing observing
conditions, or they could also originate from imperfect data reduction.
First we outline the procedure of common photometry reduction methods
along with their smooth transformations that correct for trivial
systematic variations. Then we proceed to the characterization of the
remaining systematics, and discuss them and their possible causes
through the examples of the HAT Network light curves.

The typical procedure of wide field photometry either employs  
flux extraction with aperture-photometry or Point Spread Function 
(PSF) fitting (e.g., DAOPHOT; Stetson 1987), or uses image 
subtraction (ISIS; Alard \& Lupton 1998; Alard 2000).

In the first type of approach, the instrumental magnitudes of
individual frames are transformed to a common magnitude reference
system, in order to eliminate the instrumental or observational 
changes between the epochs. For narrow field surveys, all stars 
experience the same fractional flux change, and thus the transformation 
can be treated as a constant magnitude shift with good approximation. 

For wide-field surveys, this transformation takes a more complex form, 
and for convenience it is assumed to be smooth function of the fitting 
parameters, e.g., of the co-ordinates. The transformation is established 
by iterative fitting over a substantial sample of stars and subsequent 
rejection of outlier values. The underlying assumption in this procedure 
is that CCD images of typical surveys have a large number of stars, and 
the overwhelming majority of them are non-variable above the noise-level 
(see Pojma\'nski 2003). We note in passing that other wide field surveys 
use similar methods in their basic reduction pipelines. 

Ideally, after applying the above determined transformation, the
resulting light curves should be free of systematic variations, and
constant stars should be dominated only by noise. However, omission of
some ``hidden'' parameters, characteristic of the transformation, might
lead to systematic spurious variations in the transformed magnitudes; 
sources with close values of these parameters will exhibit similar 
variations. Improper functional form of the transformation, or inclusion 
of variable stars in the fit might lead to similar systematics.
Furthermore, because local effects might dominate the light curves
(local in position or other parameters), variations remain even after
application of the best large scale (smooth) transformations. Such 
systematics are known to exist in various massive photometric surveys 
(see, e.g., Drake \& Cook 2004 for the MACHO and Kruszewski \& Semeniuk 
2003 for the OGLE projects) and even in classical CCD observations 
(Makidon et al. 2004), but in wide field observations they could be 
especially severe (Alonso et al.~2003). 

To give examples, a time-dependent PSF causes variable merging, where the
flux contribution of a star to the flux of another star and its
background annulus varies (assuming aperture photometry). The extent of
systematics induced will depend on the exact configuration of the
neighbourhood of the source. Although image subtraction convolves the
profiles of the reference frame before the subtraction so that they
have (in principle) the same width as on the frame that is analysed, 
we found that there remain small amplitude correlations with the 
periodic variation of original PSF widths. Periodic motion of stars 
across the imperfectly corrected pixels, hot-pixels or bad columns, 
or simply the gate-structure of front-illuminated pixels can be another 
cause for systematic variations. Faint stars that fall in the vertical 
vicinity of stars that are on the saturation limit might be periodically 
affected by the saturation of the bright source, e.g., with the $29^d$ 
period of moon-cycle through the increased background, or by the nightly 
variation of extinction, as the object rises and sets. 

Here and throughout the paper we employ the database on the moderately
crowded ($b=20\arcdeg$) G175 field observed by HATNet. Similar results
were obtained with other fields. Observations for G175 were made during
the fall and winter of 2003, spanning a compact interval of 202~d with
$\sim3000$ data points for each variable in $I$-band. Nearly 10000 stars
down to $I$=12 were analysed with aperture photometry from the field.
We employed a 4th order polynomial function of the x and y coordinates
to fit each frame's magnitude system to that of the reference frame.
$V$-band data were also taken on selected nights to have complementary
colour information on the sources, but this was not incorporated in the
above fits. Field G175 observations started or ended every night whenever 
the field crossed the 35\arcdeg\ horizon, and exposures were taken with 
5~min resolution with no interruption by observations of another field. 
There is obviously a (close to sidereal) daily periodicity in the 
time-base, which has 4--8~hour long sections, and 16 hour or longer gaps.
Although the field contains nearly 10000 photometered stars, we present 
tests on only the 4293 stars, among the 5000 brightest, whose light curve 
contains at least 2800 points. These bright, intensively observed and 
accurately photometered stars are the most interesting ones for shallow 
transit searches. The faintest stars in this sample have $I\sim 11.0$~mag.

After Discrete Fourier Transformation (DFT) of the light curves in the
$[0,6]$d$^{-1}$ frequency range with 20000 frequency steps, the
derived histogram of primary frequencies shows a distribution with high
peaks around $n$~d$^{-1}$ frequencies, where $n$ is a small integer. 
The histogram also shows that the 1 and 2d$^{-1}$ peaks have double 
structures, with slight offsets from the exact integer values 
(see Fig.~1). The double peak structure is attributed to the cadence 
of 1 sidereal day of the observations and its alias. Other, higher 
frequency peaks do not show double structure, only an offset, because 
they are further off from their alias counterparts leaking from the 
negative frequency regime. A very large fraction (83\%) of the peak 
frequencies fall in the $n\pm0.02$~d$^{-1}$ interval, with 
$n=$0,1,2,3,4,5. Even excluding the n=0 peak to avoid counting 
long-periodic variables as the stars showing real systematics, the 
occurrence rate of integer peak frequencies is still 69\%. Weighting 
the histogram according to the signal-to-noise ratio of the frequency 
peaks does not significantly alter its appearance.

%
%
   \begin{figure}
   \centering
   \includegraphics[width=85mm]{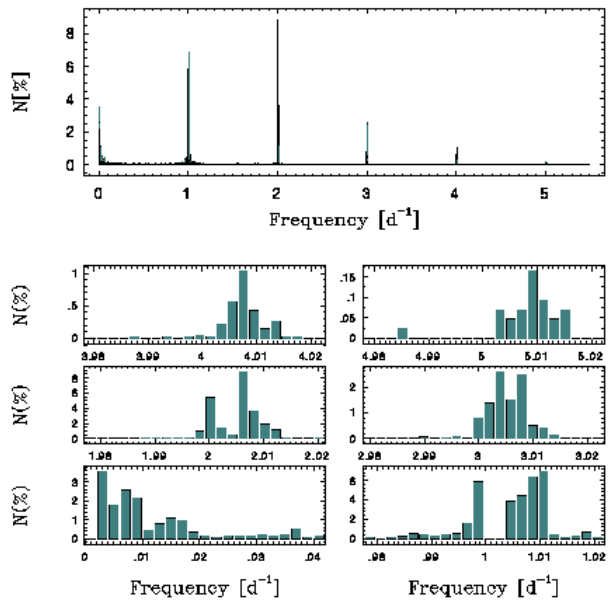}
   \caption{Distribution of the peak frequencies obtained from the DFT 
   of a sample of stars of field G175 (see text for details on the data 
   selection). {\em Largest panel:} distribution function in the full 
   bandwidth of the analysis. The sum of the bin occupation number 
   $\rm N$\% over the full frequency range is equal to 100. 
   {\em Small panels:} blow-ups of the neighbourhood of the peaks. Each 
   bin has a width of $0.002$d$^{-1}$ in all panels.           
   \label{fig:dfthist}}
   \end{figure}
%

%
   \begin{figure}
   \centering
   \includegraphics[width=85mm]{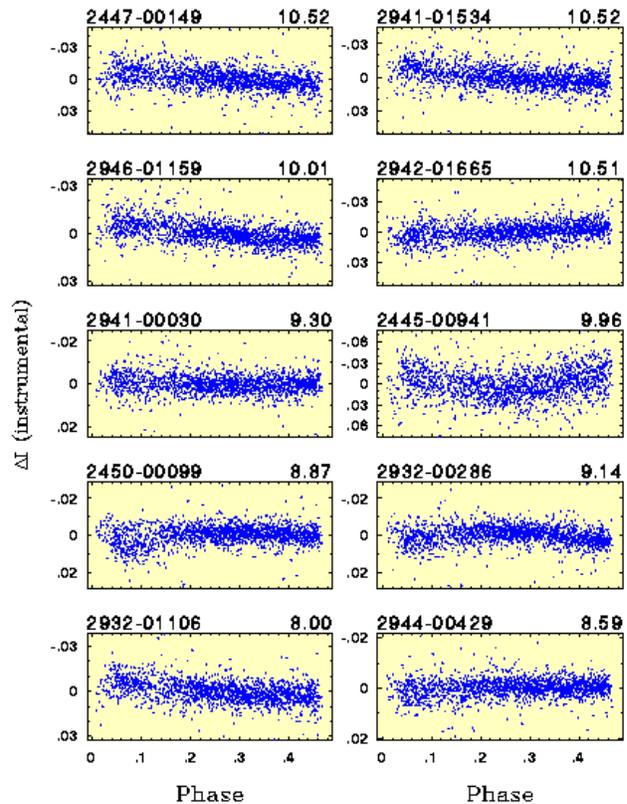}
   \caption{Examples of daily trends in field G175. The numbers on 
   the top of each panel show the star's identification (GSC number)  
   and the mean value of the instrumental magitude. Plotted are the 
   deviations from this mean value. Phases are computed with $1$~d 
   period and arbitrary epochs.  
   \label{fig:sys}}
   \end{figure}
%
%

%
%
   \begin{figure}
   \centering
   \includegraphics[width=80mm]{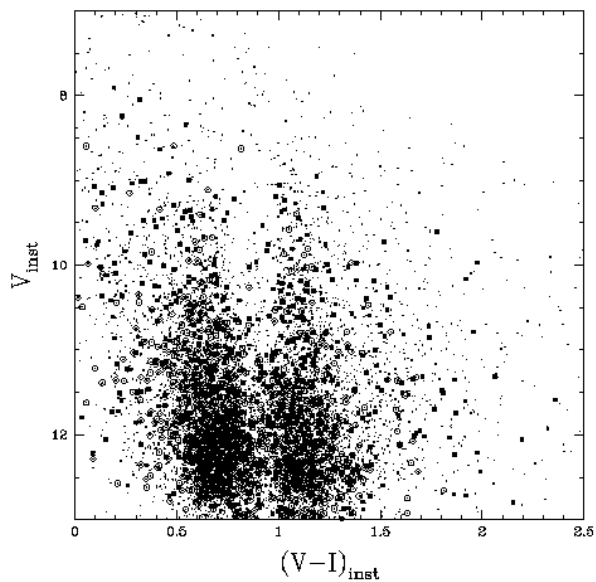}

   \caption{Colour-magnitude diagram of field G175 in instrumental V$-$I
   vs.~V system (small dots) with stars exhibiting strong 1d$^{-1}$ 
   (filled boxes) or 2d$^{-1}$ (open circles) systematics overplotted. 
   \label{fig:cmd}}
   \end{figure}
%

%
   \begin{figure}
   \centering
   \includegraphics[width=80mm]{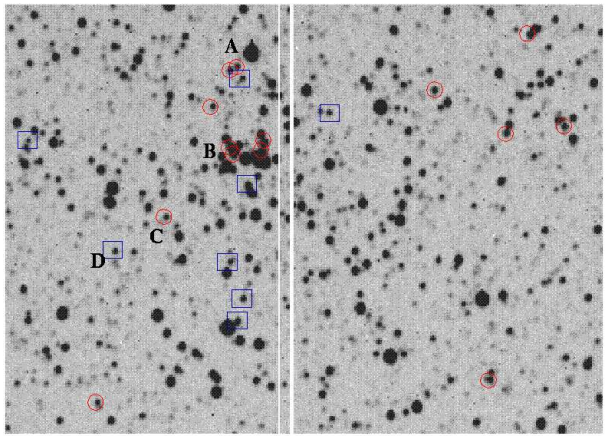}

   \caption{Image section from field G175 with dominant systematic stars
   marked by circles (1d$^{-1}$) and boxes (2d$^{-1}$). North is up, East 
   is left, image size is approx.~68\arcmin by 49\arcmin. White vertical 
   lines are due to bad CCD columns. Most of the 1d$^{-1}$ stars are 
   merging with neighbours. The bright star marked as "A" is on the 
   saturation limit, and might induce the systematic variations seen in 
   the stars South from it. The open cluster NGC2281 (marked as "B") is 
   moderately crowded with our 14\arcsec pixel-scale, and the merging 
   contributes to the 1d$^{-1}$ or 2d$^{-1}$ systematics for some of its 
   stars. Nevertheless, there are unexplained cases where isolated stars 
   also show systematic variations, such as "C" or "D".
   \label{fig:image}}
   \end{figure}

The systematic variations are of small amplitude, typically 
$\sim0.01$~mag. To have a better measure on this quantity, we derived
folded and phase-binned light curves (with 100 bins), rejected the
lowest and highest 5 points of the phased curve, and determined the
amplitude. For instance, the median value of the amplitudes of the 
1d$^{-1}$ stars is $0.015$~mag, with 85\% of them falling below 
$0.03$~mag. For illustration, in Fig.~2 we exhibit examples of daily 
trends for a few selected stars in our field.

We searched the observational parameters of HAT for commensurate
periodicities the effect of which might get superposed on the data and
cause the above mentioned systematics. The trivial daily hour-angle 
or zenith-distance variation, and the fact that we employ a
simple 4th order polynomial transformation that neglects colour
dependence suggested that colour-dependent differential extinction 
should also cause a daily periodicity: redder or bluer than average 
stars should deviate from others as the field rises or sets. Probably 
because our data are dominated by other systematics, we found no sign 
of this when plotting the distribution of higher amplitude $n$~d$^{-1}$ 
stars on a $V-I$ vs. $V$ colour-magnitude diagram (see Fig.~3). 
However, we noticed that the PSF also varies with close to 1d$^{-1}$, 
which should result in variable merging. Indeed, as shown in Fig.~4, 
the most prominent systematic variations are exhibited by stars that 
have close-by merging or almost saturated companions. On the other 
hand, there remain a few perplexing stars that seem equally merging, 
but show no strong $n$~d$^{-1}$ variations. Early image subtraction 
results on another field show slightly smaller occurrence of integer 
cycle-per-day systematics as compared to aperture photometry, which 
suggest that merging indeed plays a central role. Using more accurate 
crowded-field photometry, which takes into account the variable stellar 
profiles, the extent of systematics is somewhat suppressed. There is 
also sign for slight increase of the occurrence of systematics in the 
proximity of bad columns. In conclusion, the degree and the appearance 
of the systematics depend on many factors (position, colour, brightness, 
merging), among which merging is probably the most important in our case, 
but not the only one.


\section{Trend Filtering Algorithm (TFA)}
In this section we give a description of the algorithm in two steps. 
The first subsection outlines the method and highlights basic problems 
to be dealt with, whereas the second one reveals more technical details. 
\subsection{Preliminaries}
First of all, it is important to mention that the simplest and 
seemingly the most general approach to trend filtering would be 
the application of high-pass filtering on each daily track of 
observation. In this case with a low-order polynomial or spline 
fitting we would filter out any variation occurring on a daily 
time base, presuming that the order of the polynomial (or that 
of the spline) is properly set. Although we experimented with 
this approach at the beginning of our tests, we did not find it 
satisfactory for several reasons: (i) although the daily trends 
may be the strongest ones, other time scales play a role too; 
(ii) it is unclear what parts of the variation come from the trend 
and from the true change in the object; (iii) short tracks of 
observations may not be treated well in this way at all, because 
of the increasing chance of mixing intrinsic and systematic 
variations.

The idea of TFA is based on the observation that in a large 
photometric database there are many stars with similar systematic 
effects (this is why they are called `systematics'). As we 
discussed in the previous section, various systematics may be 
present in the individual objects. We assume that the sample size 
is large enough for allowing us to select a representative set for 
all the possible systematics. Once this subsample ({\em template set}) 
is selected, one may try to build up a filter function from these 
light curves and subtract systematics from the other, non-template 
light curves. 

The first question is how to choose the template set. Because of the 
lack of a priori information on the type of systematics influencing 
our target, selection of the template set follows only some very broad 
guidelines and constitutes basically a random set, drawn from the 
available database. More details on the template selection are given 
in Section 3.2. 

The second question that must be addressed is how to choose the filter 
function. Since we have no a priori knowledge on the functional form 
of the systematics, we take the simplest form of it, i.e., the 
{\em linear combination} of the template light curves. 

The third question is how to choose the weights of the template 
light curves in the filter function. If we assume that the variation 
in all light curves is caused only by systematics, a natural choice 
would be a simple least squares criterion for the residuals 
(observed minus filter-predicted). Since \--- as we mentioned in
Sec.~2 \--- most of the stars are non-variable, 
the criterion of minimum variance seems to be a good one for most of 
the observed stars. 

The fourth question is what distortion the TFA process causes in the 
signal being sought. It is clear that the signal will suffer from some 
distortion that can be very severe if the time scale and phase of its 
variation are close enough to those of some of the template time series. 

In view of the above considerations, TFA is designed to tackle two types 
of problems:
 
\begin{description}
\item[(a)]
{\em Increase the detection probability} by filtering out trends 
from the trend- and noise-dominated signals.
\item[(b)]
{\em Restore the signal form} by filtering out trends iteratively 
from {\em periodic} signals, assuming that the period is known. 
\end{description}

If the trends are sufficiently small (as compared to the signal 
amplitude), TFA processing is not necessary for the signal search 
(i.e., period analysis, application (a)), but (b) can still be a useful 
utility if the signal turns out to be periodic.    

The fifth question is whether the method generates unwanted signal 
components by chance inclusion of variables in the template set. In 
classical variable star observations special care was taken to avoid 
variables among the comparison stars. This was necessary to do so, 
because variable star magnitudes were obtained by simply subtracting 
the directly observed magnitudes of the variable star from those of 
the comparison stars. In the case of TFA the situation is very 
different. Here we create a linear combination of all template light 
curves which fits to the target light curve in the best way. This 
approach tends to fit features that appear in similar forms and in 
proper timings both in the target and in one or more of the template 
light curves. Because of the low likelihood this to happen for 
non-systematic variations, variables in the template set are not 
expected to influence the TFA filtering. More specifically, templates, 
containing pure sinusoidal signals will not generate detectable 
artificial signals in a target of pure white noise (see Appendix A). 
Of course, this result, referring to the statistical average of the 
signal detection parameter, does not exclude rare false signal 
detections due to statistical fluctuations.  

\subsection{Mathematical formulation}
Let us assume that all time series are sampled in the same moments and 
contain the same number of data points $N$. Let our filter be assembled 
from a subset of $M$ time series (this is the template set, the method 
of selection will be discussed later). The filter 
\{$F(i); i=1,2,...,N$\} is built up from the following linear combination 
of the template time series \{$X_j(i); i=1,2,...,N; j=1,2,...,M$\}
%
%
%
\begin{eqnarray}
F(i) & = & \sum_{j=1}^M c_j X_j(i)
\hskip 2mm .
\end{eqnarray}
It is assumed that the template time series are zero-averaged (i.e., 
$\sum_{i=1}^N X_j(i)=0$ for all $j$). The coefficients \{$c_j$\} are 
determined through the minimization of the following expression: 
%
%
%
\begin{eqnarray}
\cal D & = & 
\sum_{i=1}^N [Y(i) - A(i) - F(i)]^2 \hskip 2mm .
\end{eqnarray}
Here \{$Y(i); i=1,2,...,N$\} stands for the target time series being 
filtered, and \{$A(i); i=1,2,...,N$\} denotes the current best estimate 
of the {\em detrended} light curve, defined in the following way.

In application (a) ({\em frequency analysis}), we start from the 
assumption that the time series is dominated by systematics and noise, 
and have no a priori knowledge of any real (periodic or aperiodic) 
signal in the light curve. Consequently, in this case we set \{$A(i)$\} 
equal to the average of the target time series, i.e., 
$A(i)=\langle Y\rangle\equiv N^{-1}\sum_{k=1}^N Y(k)=const$. As we 
will see in the following sections, this simple choice of \{$A(i)$\} 
works very well, except for the rare case when the signal is similar 
to some of the templates (this happens usually for signals with long 
periods -- comparable to the length of the total time span).

In application (b) ({\em signal reconstruction}), we have already 
established (from previous analysis of the data) that the time series 
contains a periodic signal. The phase-folded time series can thus be 
used iteratively to estimate \{$A(i)$\}. First, an initial filter is 
constructed using $A(i)=\langle Y\rangle$ as above. The filtered time 
series \{$\hat Y(i) \equiv Y(i) - F(i)$\} is then phase-folded and 
binned, then {\em re-mapped} to the original time based to give a new
estimate of \{$A(i)$\}, which is in turn fed into equation~(2) to compute 
a new set of filter coefficients \{$c_j$\}. The new filter leads to a 
better determination of \{$A(i)$\}, and the iteration continues until 
some convergence criterion is satisfied. Further details and examples 
of signal reconstruction are given in Section~6.

Note that the unbiased estimate of the variance of noise of the 
filtered data can be obtained from the following equation
%
%
%
\begin{eqnarray}
\hat \sigma^2 & = & {\cal D}/(N-M) \hskip 2mm .
\end{eqnarray}
At a more technical level, the main steps of TFA are the following.

\begin{description}
\item[(1)]
Select $M$ template time series in the {\em full} field, distributed 
nearly uniformly in the field (presumably ensuring uniform sampling 
also in other parameters, e.g., in colour). Since we have no a priori 
knowledge on which stars are bona fide variables, the above selection 
is almost random, except that stars with low number of data points, 
low brightness and high standard deviation are not selected. Although 
the result is not sensitive to the actual values of the above limits, 
it is still better to employ them, in order to avoid any (however small) 
chance of biasing the target light curve. Nevertheless, for long-periodic 
target variables there is a higher chance of finding a template member 
varying on a similar time scale with similar phase. Once the template 
set is selected, it is fixed throughout the analysis.     
\item[(2)]
Define the time base to be used by the filter and target time series. 
Since in modern automated surveys nearly all photometered stars have 
the same number of data points distributed in the same moments of time, 
selection of the uniform time base is made on a subsample of the 
template light curves (containing $\sim 50$ stars in our case). 
We select the time base from the template light curve that contains 
the largest number of data points. Occasionally, data points, at some 
moments defined by the template time base, might be missing in the 
observed light curve. In these cases they are filled in with the 
average value of that light curve.
\item[(3)]
Compute zero-average template time series by using some criterion for 
outlier selection (in our case a 5$\sigma$ clipping).  
\item[(4)]
Compute normal matrix from the above template time series:
%
%
%
\begin{eqnarray}
g_{j,k} = \sum_{i=1}^N X_j(i)X_k(i) \hskip 1mm; 
\hskip 5mm j,k=1,2,...,M\hskip 2mm ,
\end{eqnarray}
and compute the inverse of it: \{$G_{j,k}$\}.
\item[(5)]
{\em For each light curve}, 
compute scalar products of the target and template time series:
%
%
%
\begin{eqnarray}
h_j = \sum_{i=1}^N \tilde Y(i)X_j(i)\hskip 2mm .
\end{eqnarray}
Here the modified target time series \{$\tilde Y(i)\equiv Y(i)-A(i)$\}
is also assumed to be free of outliers.
\item[(6)]
Compute solution for \{$c_j$\} and apply correction accordingly: 
%
%
%
\begin{eqnarray}
c_j = \sum_{k=1}^M G_{j,k}h_k\hskip 2mm ,
\end{eqnarray}
The corrected time series is computed from 
%
%
%
\begin{eqnarray}
\hat Y(i) = Y(i)-\sum_{k=1}^M c_jX_j(i)\hskip 2mm ,
\end{eqnarray}
\end{description}
It is important to make the following comments concerning 
additional details of the computational implementation.

A significant advantage of the above algorithm is that there is no 
need to compute the normal matrix for each target separately. This 
means only the computation of simple array-array and matrix--array 
products (steps 5--6 above). For period search this is done only 
once for each star, but for signal reconstruction this is repeated 
several times (which is still not too time-consuming). 

Since the template set is fixed, in principle the normal matrix need 
only be computed once. In practice, if the target accidentally coincides 
with one of the template components, the latter must be eliminated from 
the normal matrix before computing the filter. This requires a reshuffling 
and inverting of the normal matrix, but it should happen fairly rarely, 
because of the relative low number of template time series 
($\sim 10^2 \-- 10^3$) compared to the size of database 
($\sim 10^4$ stars). The CPU request for the initial setup of a 
filter with few hundred templates is only a few times slower than 
a single run of the BLS algorithm for transit search. Nevertheless, 
for large template numbers the extra matrix manipulations increase 
the execution time substantially. For example, on a 3GHz commercial PC 
a full BLS analysis of a set of 13000 time series with 2300 data points 
per time series, 30000 frequency steps per time series and 920 TFA 
templates required 100 CPU hours. The cost of the non-TFA analysis of 
the same dataset was only one fifth of it.

The computer memory required by TFA can be quite large. This is mostly 
because of the template set, since it requires the usage of $N\times M$ 
floating point array elements. Inversion of the normal matrix may pose 
also some numerical problems if $M$ is too large, although we have not 
encountered such problems even for very large $M$ such as $800$. 
This stability is probably due to the basically uncorrelated noise 
of the the light curves.   

We note that a possible way to decrease the number of templates is to 
perform a Singular Value Decomposition (SVD, see Press et al. 1992) 
and employ only the `significant' eigenvectors derived from this 
decomposition. Although we experimented with SVD at some point, we 
decided that a clearcut division of the eigenvectors into `significant' 
and `non-significant' ones was not possible. Therefore, the original 
`brute force', full template set approach was followed.

There is also the question if magnitude or intensity (flux) values 
should be used in the above procedure. We experimented with both 
quantities and finally settled at the direct magnitude values, since 
no definite advantage was observed to result from a conversion to fluxes.


\section{Trend suppression tests}
The simplest way to check the most straightforward effect of TFA 
is to compare the standard deviations of the original and TFA-processed 
time series. This comparison, of course, will not tell us if all temporal 
distortions due to trends have been successfully filtered out or not. 
Nevertheless, it will at least give us some indication on the efficiency 
of the method. 

The effect of decreasing the standard deviation is displayed in 
Fig.~5. Unbiased estimates of the standard deviations $\sigma$ 
for the TFA runs were computed with the aid of equation~(3) by 
using a template number of $M=361$. (Originally we set $M=400$, but 
limitations on the template time series \--- see Section~3.2 \--- 
lowered this value. A similar statement is applied to other template 
numbers used in this paper.) We see that with the increase 
of the standard deviation $\sigma$(non-TFA) of the original time 
series, TFA is likely to introduce significant corrections. This, as 
expected, means that toward larger $\sigma$(non-TFA) (i.e., at 
fainter magnitudes) there are time series that are more affected 
by systematic errors than most of the brighter stars. There are 
relatively sharp upper and lower boundaries in the decrease of  
the standard deviation. The existence of the upper boundary at small  
corrections suggests that {\em all} stars are affected by some 
(however small) systematic errors and TFA is capable of filtering 
out a substantial part of them. Although TFA also has some side-effects 
on any real signal in the time series, it suppresses the systematics 
much more effectively than the signal, as shown by the fact that it 
leads to a significant increase in detection probability (see Section~5).

%
%
   \begin{figure}
   \centering
   \includegraphics[width=80mm]{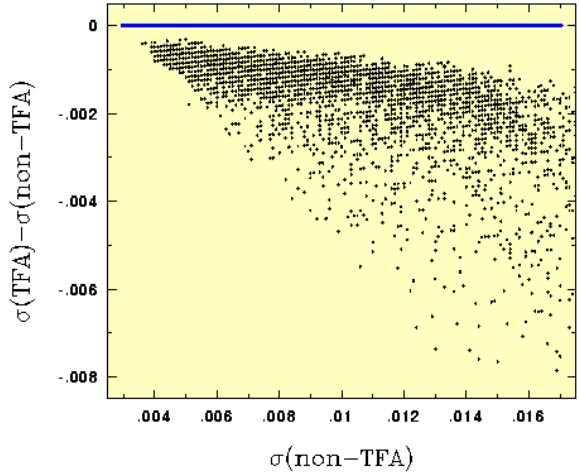}
      \caption{Decrease of the unbiased estimate of the standard 
      deviation of the time series due to the application of TFA.  
      The brightest 4293 stars with greater than 2800 data points 
      per star of field G175 are used. The TFA result was obtained 
      with 361 templates. For better visibility of the minimum 
      decrease in $\sigma$, we plotted a horizontal line to indicate 
      the zero correction level.    
}
         \label{}
   \end{figure}
%
%
In order to examine the question whether the decrease of the standard 
deviation has also led to the diminishing of the temporal signatures 
of the systematics, we frequency analysed the original and 
TFA-processed light curves. The BLS routine was run on both datasets. 
The analysis was performed in the [0.01,0.99]d$^{-1}$ 
band that covers the period of interest in transit search, but excludes 
very  long periods close to the total time span. We recall that 
the BLS routine generates several aliases (subharmonics) from the 
daily trends at frequencies of small integer ratios (see K02). This 
effect results in the appearance of several aliases in the frequency 
band chosen for this test. Therefore, the success of TFA in filtering 
out systematic periodicities can be judged from the above analysis. 

The distribution function of the frequencies of the highest peaks 
obtained on the original, unfiltered dataset is shown in the upper 
panel of Fig.~6. The large number of stars affected by the daily 
trends (that appears here mostly at $\sim 0.5$d$^{-1}$) is somewhat 
surprising. Actually, it turns out that 24\% of the highest 
peaks fall in the [0.49,0.51]d$^{-1}$ frequency range and about 
22\% of them appear in the $\pm 0.01$d$^{-1}$ neighbourhood of 
0.02, 0.33 and 0.66d$^{-1}$. It is clear that a simple direct 
use of the data for transit search would leave a quite substantial 
part of the sample not utilized because of the domination of trends. 

Next, the above analysis is repeated by applying TFA. The result is 
displayed in the lower panel of Fig.~6. We see that severe aliases 
due to the daily trends disappeared. Now, as it can be expected for 
a sample with time series of mostly pure noise, all frequency bins 
are nearly uniformly occupied. The remaining fluctuations are all of 
statistical origin, due to the low number of events falling in the 
narrow bins. 

We note in passing that we performed the above test also on the peak 
frequencies based on DFT spectra. Although at low or moderate template 
numbers we still got some remnants under the 1\% level in the frequency 
distribution around integer frequencies, at high number of templates 
we got basically a flat distribution at the 0.3\% level, very similar  
to the one we got for the BLS spectra. We recall that 69\% of the DFT 
peak frequencies obtained by the non-TFA DFT analysis fell in the 
$\pm 0.02$~d$^{-1}$ neighbourhood of positive integer frequencies.      

%
%
   \begin{figure}
   \centering
   \includegraphics[width=85mm]{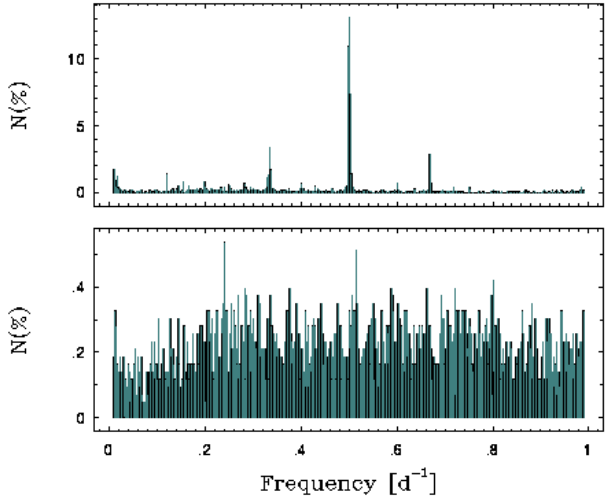}
      \caption{Comparison of the distribution functions of the peak 
       frequencies obtained in the BLS analysis with and without 
       TFA processing ({\em lower} and {\em upper} panels, respectively). 
       Each bin has a width of $0.002$d$^{-1}$. The sum of the bin 
       occupation number $\rm N$\% over the full frequency range is 
       equal to 100. The same number of data points and templates are 
       used as in Fig.~5.        
}
         \label{}
   \end{figure}
%


\section{Transit detection tests}
We examine the periodic transit detection capabilities of the BLS 
method on the TFA-processed light curves. This is done with the 
aid of various test signals generated from the observed time series 
and synthetic signals. To each of the 4293 observed, zero-averaged 
light curve we add the transit signals given in Table~1. Although 
most of the tests presented in this paper refer to signal \#1, 
the other test signals were also utilized in order to check the 
efficiency of the method for various signal parameters. Most of 
these parameters correspond to realistic transit configurations, 
except perhaps for \#5, where the relative duration of the transit 
is somewhat too long.    

In order to illustrate the signal detection capability after applying 
TFA, in Fig.~7 we exhibit two specifically chosen examples for the 
extreme improvement one can get for trend-dominated signals. It is 
obvious that TFA is capable of suppressing a considerable part of 
the trend without a simultaneous suppression of the periodic signal 
component. Although the signal also suffers from some amount of 
suppression, in a very large number of cases this is smaller than 
the one exerted on the trends. This, as we will see below, ultimately 
leads to significantly higher detection rates for TFA-processed signals.  

%
%
\begin{table}
\caption[ ]{Parameters of the synthetic signals}
\begin{center}
\begin{tabular}{ccccc}
\hline
No. & $P$[d] & $\delta$[mag] & $\varphi$ & $q$ \\
\hline
\#1 & $5.12345$ & $-0.015$ & $0.73$ & $0.03$ \\
\#2 & $1.23456$ & $-0.015$ & $0.33$ & $0.03$ \\
\#3 & $1.23456$ & $-0.005$ & $0.33$ & $0.03$ \\
\#4 & $1.49000$ & $-0.015$ & $0.33$ & $0.03$ \\
\#5 & $12.5000$ & $-0.015$ & $0.33$ & $0.03$ \\
\hline
\end{tabular}
\end{center}
{\footnotesize
\underline {Comments:}
\begin{description}
\item[$P$]= Period;
\item[$\delta$]= Depth of the transit;
\item[$\varphi$]= Phase of the transit, defined as $(t_{in}-t(1))/P$, 
where $t_{in}$ is the moment of the first ingress after $t(1)$, the 
time of the first item of the time series;
\item[$q$]= Relative length of the transit, defined as the ratio of 
the actual transit length and the period.
\end{description}
}
\end{table}
%

%
%
   \begin{figure}
   \centering
   \includegraphics[width=85mm]{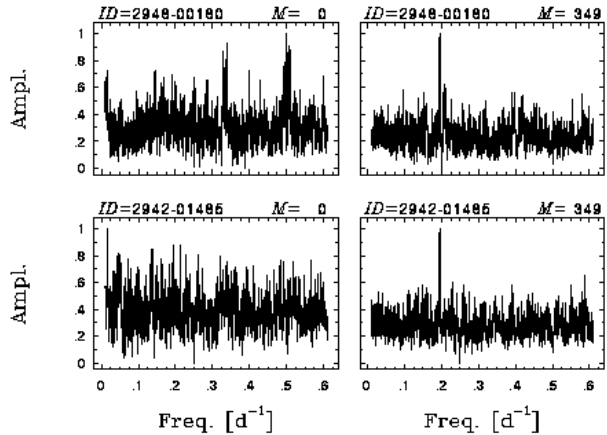}
      \caption{Examples of the signal detection capability of TFA. 
      In both cases test signal \#1 was added to the observed light 
      curves of the stars given by the identification numbers in 
      the headers.{\em Left column:} BLS spectra of the non-TFA 
      (original) test light curves. {\em Right column:} BLS spectra 
      of the TFA test light curves. Template numbers are shown in 
      the headers. The spectra are normalized to unity in each panel 
      separately. 
}
         \label{}
   \end{figure}
%

In order to compare the detection rates obtained from the original 
and TFA-processed light curves, we need to define the meaning of 
the word `detection'. Based on the frequency spectrum, we employ the 
following two criteria to regard a peak frequency as being identical 
with the injected test frequency.

\begin{description}
\item[$-$]
The highest peak in the BLS spectrum should have a frequency in the 
$[f_{\rm test}-0.001,f_{\rm test}+0.001]$d$^{-1}$ interval, where 
$f_{\rm test}$ is the frequency of the test signal, in the present 
case of signal \#1 it is equal to 0.1952d$^{-1}$.
\item[$-$]
The Signal to Noise Ratio (SNR) of the highest peak in the frequency 
spectrum must be greater than 6. The definition of SNR is the following: 
${\rm SNR} = [p({\rm highest\hskip 2mm peak})-\langle p\rangle ]/\sigma(p)$. 
Here $p$ denotes the BLS statistics (denoted by SR in K02), 
$\langle p\rangle$ is the average power in the frequency band analysed, 
$\sigma(p)$ is the standard deviation of the spectrum (we omit large 
peaks in the computation of $\langle p\rangle $ and $\sigma(p)$). 
\end{description}

The above lower limit of SNR is somewhat arbitrary and it is a 
compromise between avoiding too high false alarm rates but not losing 
too many positive peak identifications. At the end of this section 
we will discuss the effect of posing a more stringent condition on 
SNR. 
 
For the  characterization of the detection probability, we introduce 
the Cumulative Detection Probability (CDP) that is defined in the 
following way:
%
%
\begin{eqnarray}
{\rm CDP} & = & {N_d(m<I)\over N_{nv}(m<I)}
\hskip 2mm .
\end{eqnarray}
Here $N_d(m<I)$ denotes the number of detections for test signals
brighter than $I$ magnitudes, $N_{nv}(m<I)$ stands for the number 
of non-variable stars in this magnitude range. Because some 5--10\% 
of the stars in the sample are true variables and they will either dominate 
the signal or decrease the SNR of the test signal below the detection 
limit, conservatively we take $N_{nv}(m<I)=0.95N_t(m<I)$, where 
$N_t(m<I)$ is the total number of stars brighter than $I$ magnitude. 
Because in this test {\em all} stars have injected transit signals, 
the detection rates to be derived have nothing to do with the true 
incidence rates of transits in our sample or in stars in general. 
The sole purpose of this test is to find out by how much do we 
increase the chance of detection by applying TFA.  

The derived CDPs on the sample of 4293 stars are shown in Fig.~8. 
The somewhat jagged variation of the curves for bright stars is 
attributed to the relatively small sample size for these stars. 

%
%
   \begin{figure}
   \centering
   \includegraphics[width=80mm]{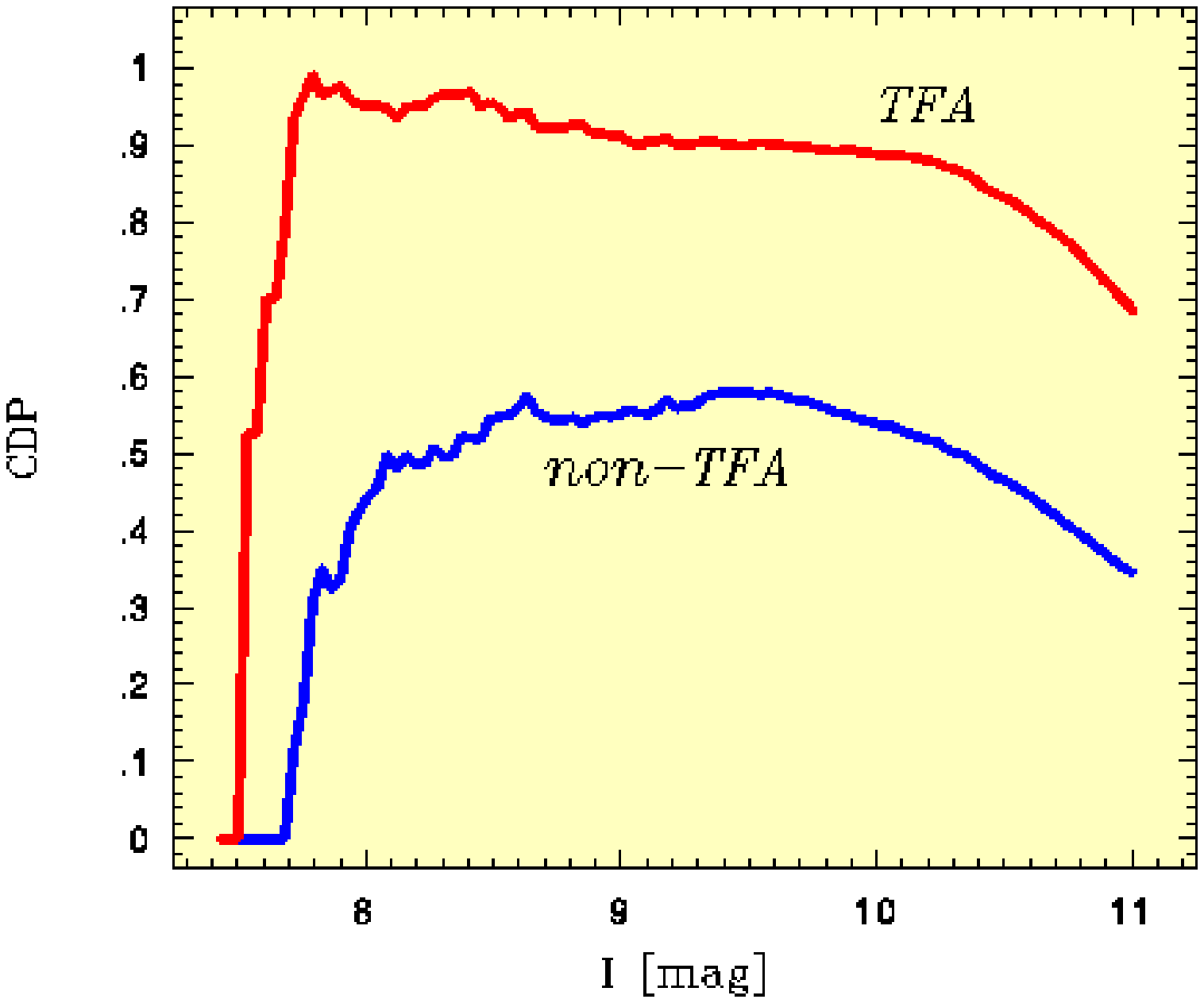}
      \caption{Cumulative Detection Probability for test signal \#1 in 
      the presence of observational noise. All cases considered have 
      SNR$>6$. The TFA result was obtained with 595 templates. 
}
         \label{}
   \end{figure}
%

It is clear from this figure that TFA introduces a substantial 
improvement in our detection capability. The high and almost constant 
detection rate down to $\sim 10.2$~mag indicates the steady performance 
of TFA. The situation changes at fainter magnitudes, when the random 
photometric noise becomes higher. This, coupled with the not too 
favorable combination of period, phase and data distribution for signal 
\#1, results in a definite decrease in CDP at fainter magnitudes. 
The decrease of the detection probability is more visible if we check 
narrow magnitude intervals. For example, for test signal \#1, the 
detection probability decreases from $0.45$ to $0.08$ for the raw data 
when moving from the $\pm 0.1$ range of $10.0$~mag to the same range 
of $11.0$~mag. The same probabilities for the TFA-processed light curves 
are $0.87$ and $0.27$, respectively. For short period signals the 
detection capability (both for the original and TFA-processed time 
series) is greater than for long period signals as is to be exhibited 
at the end of this section.

%
%
   \begin{figure}
   \centering
   \includegraphics[width=80mm]{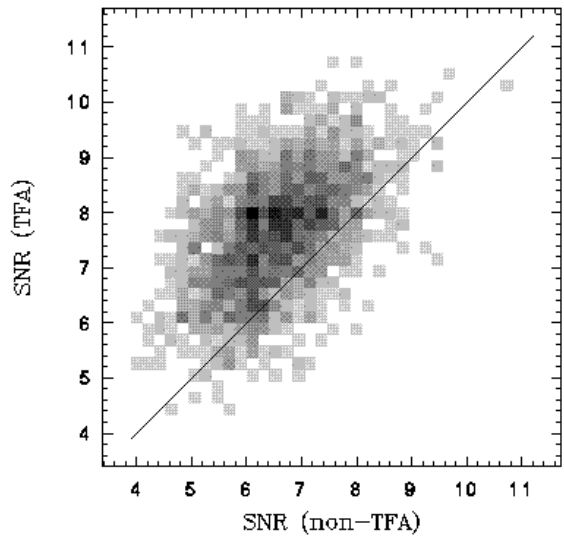}
      \caption{Comparison of the SNRs obtained for test signals \#1 
      with BLS spectra satisfying only the frequency constraint for 
      detectability. Shading is proportional to the number 
      of cases (darker regions contain larger number of cases). 
      The TFA result was obtained with 595 templates.}
         \label{}
   \end{figure}
%

It is also important to characterize the change in the SNR when we lift 
the SNR$>6$ constraint. Then, by keeping only the frequency constraint 
given above, and plotting the SNR values for the variables satisfying 
this single constraint (both in the TFA-processed and original datasets), 
we get the plot displayed in Fig.~9. Again, the improvement introduced 
by TFA is obvious. Many transits, with SNR$<6$ in the original time 
series become confidently `discovered' after applying TFA. However, 
it is also seen that there are cases when the original time series 
have higher SNR. Although this feature is expected to be rare, it is 
not too surprising. While TFA suppresses mostly those features in the 
target light curve that can also be found in the template set, it may 
happen that during this procedure, features (i.e., intrinsic variability) 
not representative of the trends, become also suppressed. This effect 
of TFA is not too harmful at higher SNR values (e.g., for SNR$>8$), 
because the signal will be securely detected in both datasets, albeit 
with a somewhat smaller significance for the TFA-processed time series. 
The situation becomes worse at lower SNR, when the signal may become 
suppressed under the detection limit. 

In order to get a more quantitative insight into this phenomenon, 
we checked the number of cases when the signal was detected in the 
original, but escaped detection in the TFA-processed time series. 
By performing this and the opposite check with two different template 
numbers and detection limits for various test signals, we obtained 
the results shown in Table~2. It is seen that the number of such exclusive 
detections is always much higher among the TFA-processed light curves. 
In addition, the application of higher template numbers increases the 
number of cases when only TFA is capable to detect the signal. 
Nevertheless, it seems that there is almost always a relatively small 
fraction of stars that escape detection in TFA, because of the unwanted 
side effect of trend suppression.

%
%
\begin{table}
\caption[ ]{Mutually exclusive detections for various test signals}
\begin{center}
\begin{tabular}{crrrr}
\hline
No. & $N_6^{361}$ & $N_6^{820}$ 
    & $N_8^{361}$ & $N_8^{820}$ \\
\hline
\#1 &   56 &   59 &   29 &   21 \\
    & 1414 & 1696 &  692 & 1452 \\
\#2 &    0 &    1 &    0 &    0 \\
    &  619 &  704 &  845 &  930 \\
\#3 &   24 &   28 &    5 &    8 \\
    & 1014 & 1139 &  838 &  980 \\
\#4 &   56 &   82 &   37 &   69 \\
    & 1162 & 1264 & 1176 & 1323 \\
\#5 &    0 &   19 &    0 &    0 \\
    &    0 & 1957 &    0 &  177 \\
\hline
\end{tabular}
\end{center}
{\footnotesize
\underline {Comments:}
\begin{description}
\item[$-$]
$N_i^j=$~Number of exclusive detections obtained with SNR$>i$
and template number $j$;
\item[$-$]
{\em Upper rows:} not detected using TFA-processed data, but detected 
using raw data
\item[$-$]
{\em Lower rows:} detected using TFA-processed data, but not detected 
using raw data
\end{description}
}
\end{table}
%

It is interesting to compare the effect of various signal parameters 
on the detection probability. As before, here we limit ourselves to 
the test signals given in Table~1. For each test signal we check the 
CDP obtained at two magnitude and at two SNR levels entering in the 
detection condition discussed above. Changing the SNR level is 
illustrative for the expected decrease of CDP when one is aimed at  
detections with high confidence and with practically zero false alarm 
rate. The derived CDP values are given in Table~3. As expected, TFA 
always yields {\em higher} CDP, independent of the signal. The gain  
is of course smaller for cases of high signal-to-noise ratio, such as 
for signal \#2. It is noticeable that single site observations inevitably 
lead to bias toward discovering short periodic transits 
(see also Brown 2003). For example, test signals \#2 and \#5 have the 
same parameters except for their periods. This results in a dramatic 
difference in CDP. In terms of the signal-to-noise ratio of the frequency 
spectra, SNR rarely hits 8 even when TFA is employed for \#5, whereas 
an overwhelming majority of the spectra of the original (non-TFA) data 
for \#2 have SNR$>8$ and quite often above 10 (reaching a maximum value 
of 25).

%
%
\begin{table}
\caption[ ]{CDP for various test signals}
\begin{center}
\begin{tabular}{ccccc}
\hline
No. & $CDP_6^{10}$ & $CDP_6^{11}$ & $CDP_8^{10}$ & $CDP_8^{11}$ \\
\hline
\#1 & 0.54 & 0.35 & 0.05 & 0.03 \\
    & 0.91 & 0.73 & 0.64 & 0.38 \\
\#2 & 0.90 & 0.83 & 0.86 & 0.76 \\
    & 1.00 & 1.00 & 0.99 & 0.98 \\
\#3 & 0.36 & 0.17 & 0.19 & 0.08 \\
    & 0.79 & 0.44 & 0.68 & 0.32 \\
\#4 & 0.72 & 0.51 & 0.62 & 0.37 \\
    & 0.95 & 0.79 & 0.92 & 0.67 \\
\#5 & 0.05 & 0.03 & 0.00 & 0.00 \\
    & 0.73 & 0.50 & 0.08 & 0.04 \\
\hline
\end{tabular}
\end{center}
{\footnotesize
\underline {Comments:}
\begin{description}
\item[$-$]
$CDP_i^j=$~CDP at magnitude $j$ for stars with SNR$>i$;
\item[$-$]
{\em Upper rows:} non-TFA results, {\em lower rows:} TFA results with 
820 templates.
\end{description}
}
\end{table}
%


\section{Signal reconstruction}
In Section~3 we emphasized that when using TFA for signal detection, 
we assume that the signal is trend- and noise-dominated, therefore 
the use of constant detrended signal \{$A(i)=$const\} seems to be 
justified. Nevertheless, the original signal will suffer from some 
level of distortion, because of the requirement of minimum variance 
for the signal that is assumed to be constant. It is clear that 
for non-periodic signals of arbitrary shape, sampled unevenly and 
gapped in time, is very difficult to make any reasonable initial 
guess on the original signal form. Therefore, the reconstruction 
method developed in this section can be applied only to {\em periodic} 
time series. 

The main step of the algorithm is the iterative approximation of    
\{$A(i)$\}. This has already been described in Section~3. It is 
important to find a good method to estimate \{$A(i)$\} at each 
step of iteration. Here we employ the simple method of bin averaging 
to derive the updated set of \{$A(i)$\} at each step of iteration. 
Of course, the reliable estimate of \{$A(i)$\} requires the use of 
proper number of bins in order both to allow an ample sampling of 
the light curve and to ensure statistical stability due to observational 
noise. Although in our case of $\sim 3000$ data points per light curve 
even $100$ bins give a reasonable noise averaging, usually lower number 
of bins (e.g., 50) are also acceptable, because of the smooth shapes 
of most of the light curves. 

Once the signal shape is more accurately determined by the above 
general method, one may proceed further and utilize this information 
to derive a more specific model for the signal shape. For example, if 
the signal turns out to be of box-shaped, one can use the BLS routine 
to get more accurate approximation for \{$A(i)$\}. However, it is 
emphasized that this second level of filtering can be used only if 
the first, more general method firmly supports our assumption on the 
signal form. Otherwise the output will be biased through our incorrect 
assumption on the signal shape. 

Iteration on the folded light curve continues until the relative difference 
between the standard deviations of the residuals (see equations~(2) and (3)) 
in the successive iterations become under a certain limit. We set this 
limit at $10^{-3}$.
 
%
%
   \begin{figure}
   \centering
   \includegraphics[width=85mm]{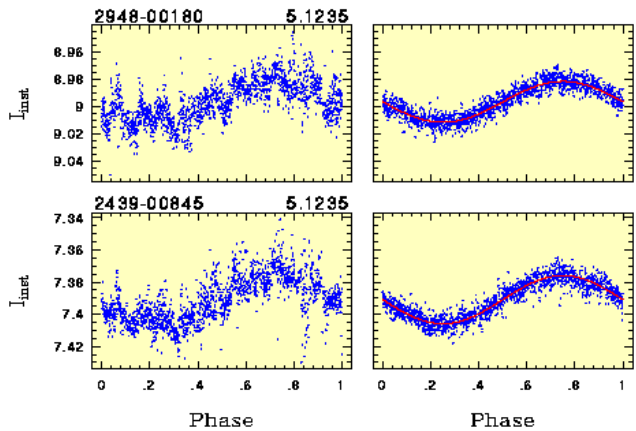}
      \caption{Reconstruction of a sinusoidal {\em test signal} with 
      the aid of iterative TFA filtering. {\em Left column:} original 
      folded signal. {\em Right column:} TFA-filtered signal with the 
      synthetic signal shown by continuous line. Identifiers (GSC numbers) 
      of the stars employed in the generation of the test signals and 
      the corresponding periods are given in the headers of the panels 
      on the left. We used 595 template light curves and bin averaging 
      in the TFA filtering. 
      }
         \label{}
   \end{figure}
%

%
%
%
   \begin{figure}
   \centering
   \includegraphics[width=85mm]{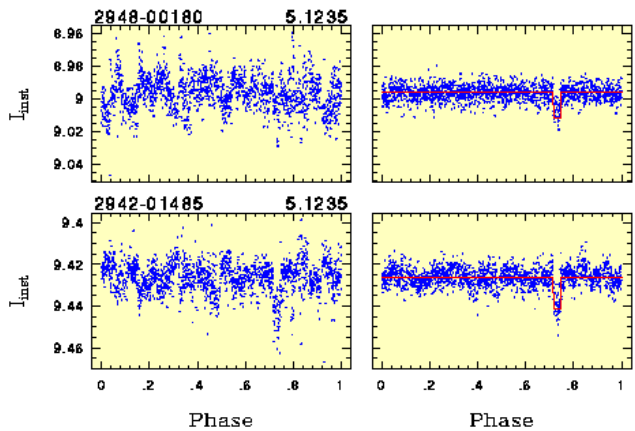}
      \caption{Reconstruction of our standard box-shaped test signal 
      \#1 (see Table~1) with the aid of iterative TFA filtering. 
      Notation, method of reconstruction and the TFA parameters used 
      are the same as in Fig.~10. 
      }
         \label{}
   \end{figure}
%

%
%
%
   \begin{figure}
   \centering
   \includegraphics[width=85mm]{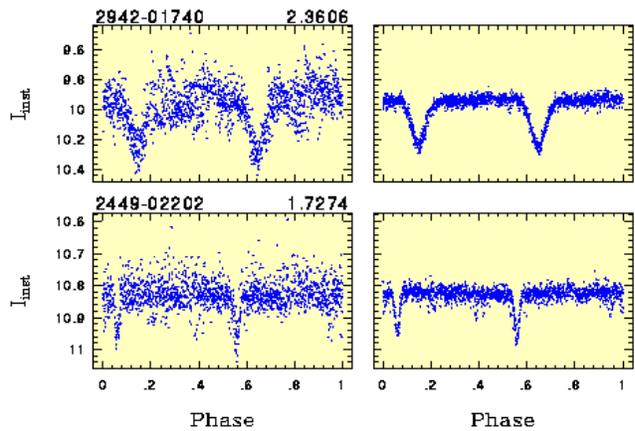}
      \caption{Examples of the reconstruction of the light curves of 
      {\em observed} intrinsic variables with the aid of iterative TFA 
      filtering. Notation, method of reconstruction and the TFA parameters 
      used are the same as in Fig.~10. 
      }
         \label{}
   \end{figure}
%

For the illustration of the efficiency of the method, first we generate 
a sinusoidal test signal in the same way we did in the case of the 
transit signals in Section~5 (i.e., by injecting the synthetic signal in 
the real light curves). The period of the injected signal is the 
same as that of test signal \#1 (see Table~1), whereas its amplitude 
is $0.015$~mag. We use 100 bins for the estimation of the detrended  
light curve \{$A(i)$\} during each step of iteration. In Fig.~10 we 
show two examples of the substantial improvement obtained after TFA 
filtering. As we see, the TFA-filtering returned the original signal 
form without suppressing the amplitude or modifying the phase. 

The second example in Fig.~11 shows the result for our standard box-shaped 
transit signal (\#1, see Table~1). Again, it is seen that the signal can 
be completely scrambled by systematics, yet TFA is able to reconstruct 
the original signal shape (and also to find the correct period). 

In order to quantify the accuracy of the estimated parameters when 
TFA signal reconstruction is used, we compare the derived transit 
parameters with those obtained in the course of the BLS period search 
(i.e., without iterative TFA correction of the transit shape). 
In Table~4 we show the averages and standard deviations of the three 
parameters (computed only from the significant detections, as defined 
in Section~5). We see that both the phase and the width are estimated  
with the same accuracy in both cases, with no apparent systematic 
errors. However, the depth is substantially underestimated if TFA 
is applied without iterative signal reconstruction.

%
%
\begin{table}
\caption[ ]{Estimated parameters of test signal \#1}
\begin{center}
\begin{tabular}{lrc}
\hline
Parameter & Average & $\sigma$ \\
\hline
$\delta$  & $-0.0111$ & 0.0013 \\
          & $-0.0147$ & 0.0016 \\
q         &   0.0290  & 0.0019 \\
          &   0.0290  & 0.0019 \\
$\varphi$ &   0.7307  & 0.0043 \\
          &   0.7308  & 0.0043 \\
\hline
\end{tabular}
\end{center}
{\footnotesize
\underline {Comments:}
\footnotesize{
\begin{description}
\item[$-$]
Transit parameters are described in Table~1.
\item[$-$]
Only significant detections (as defined in Section~5) have been considered.
\item[$-$] 
At each parameter the upper row refers to the values obtained {\em without}, 
whereas the lower one {\em with} iterative TFA filtering.
\item[$-$]
$\sigma$ denotes the standard deviations of the detected signal parameters.
\end{description}   
}}
\end{table}
%

Finally, in Fig.~12 we show two real examples for signal reconstruction 
from the G175 database. As in all of our examples presented in this section, 
we used bin averaging applicable for arbitrary-shaped signals. Although 
these examples are not representative of the overall effect of TFA on the 
other variables in the database, they clearly show the size of improvement 
one may get in cases heavily corrupted by systematics.


\section{Conclusions}
Current studies indicate that the incidence rate of hot Jupiters is 
much lower than it was believed a few years ago (Brown 2003). Although 
we do not have good observational constraint yet for bright Galactic 
field stars, based on the OGLE survey of faint Galactic Bulge 
stars and the very low number of positive cases obtained by 
spectroscopic follow-ups (Alonso et al. 2004; Pont et al. 2004, and 
references therein), the predicted low incidence rate could be close 
to reality. Therefore, it is of prime importance to develop effective 
methods both for data reduction and for subsequent data analysis. 
The goal to be reached by photometric surveys targeting extrasolar 
planets is twofold: 
(i) to reach photometric accuracy close to the photon noise; 
(ii) to minimize systematic effects leading to coloured noise and 
therefore seriously jeopardizing the discovery rate. Although current 
sophisticated data reduction techniques (e.g., Differential Image 
Analysis by Alard \& Lupton 1998; Alard 2000) may help to reach the near 
photon noise limit and minimize systematic effects, the latter seems 
to be never eliminated completely. Systematic effects, usually appearing 
on a daily timebase, are present even in observations that cover small 
area of the sky per frame (MACHO, OGLE). Wide field surveys 
have severe additional disadvantages (e.g., narrow PSF, strong 
nonlinearities in the magnitude/position transformations, etc.) 
that amplify further the systematic effects and thereby make subsequent 
data analysis more difficult. 

Since the appearance of systematics seems to be generic, it is important 
to devise a method that is capable of filtering out systematics from 
the time series photometry in a post-reduction phase. The reason why 
such an approach may work is that in standard data reduction pipelines 
the photometry is done through image processing of snapshots. Therefore, 
these types of data reduction methods are unable to consider the time 
series properties of the database. The Trend Filtering Algorithm (TFA) 
described in this paper takes into consideration the temporal behavior 
of the data by constructing a linear filter from a template set of time 
series chosen from the database to be analysed. Our TFA is capable to 
handle two problems:

\begin{description}
\item[(a)]
Create optimally filtered time series for frequency analysis; 
\item[(b)]
Reconstruct periodic signals scrambled by systematics.
\end{description}  

In both cases the optimization of the filter is made in the standard 
least squares sense. In application (a) the signal is assumed to be 
trend- and noise-dominated, whereas in (b) this assumption is not made 
and the filtered signal is iteratively built up by applying case (a) 
assumption on the residual (observed minus cycle-averaged) time series. 

Tests made on the database of the HATNet project have shown that by 
the application of TFA the signal detection rate (the Cumulative 
Detection Probability -- CDP) increases by up to 0.4 for stars brighter 
than 11~mag. The improvement in the Signal to Noise Ratio (SNR) 
is often spectacular, leading to secure discoveries from signals 
originally completely hidden in the systematics. Once the period is 
found, the signal can be reconstructed by applying TFA iteratively 
as mentioned above. 

Results presented in this paper indicate that there is a steady 
increase in CDP and SNR even for template numbers above 500. This 
suggests that systematics show up in many different forms and one 
needs to consider as many of them as possible when targeting the 
detection of weak signal components. Limitations to very high template 
numbers are set only by the size of the sample, the number of the 
data points, statistical and numerical stability and computational 
power.

\section*{Acknowledgments}
A part of this work was done during  GK's stay at CfA. He thanks for 
the support and hospitality provided by the staff of the Institute. 
Thorough review of the paper by the referee is greatly appreciated.  
We thank to Kris Stanek for the useful discussions. Operation of 
the HATNet at the Fred Lawrence Whipple Observatory, Arizona, and the 
Submillimeter Array, Hawaii has been supported by the Harvard-Smithsonian 
Center for Astrophysics. We are grateful to Carl Akerlof and the ROTSE 
project (University of Michigan) for the long-term loan of the lenses 
and CCDs used in HATNet. Supports from NASA grant NAG5--10854 and OTKA 
grant T--038437 are acknowledged.


\appendix

\section{}

The purpose of this Appendix is to show that templates containing 
periodic signals do not induce similar components at the level of 
detection in a pure noise target time series. For simplicity we 
discuss only the case for a single template. The result holds also 
for arbitrary number of orthogonal templates.

Let the target \{$y_i; i=1,2,...,n$\} be equal to a pure Gaussian 
white noise series \{$\eta_i$\} with the following properties
\begin{eqnarray}
E(\eta_i) = 0  \hskip 5mm   E(\eta_i\eta_j)=\delta_{ij}\sigma^2 
\hskip 2mm ,
\end{eqnarray}
where $E(.)$ denotes the expectation value, $\delta_{ij}$ is the 
Kronecker delta function, $\sigma$ is the standard deviation. The 
template is a simple harmonic function in the form of
\begin{eqnarray}
x_i = A\sin\Phi_i\hskip 2mm ,
\end{eqnarray}
where $\Phi_i=\omega_0 t_i + \varphi_0$. The TFA minimization criterion 
leads to the following expression for the estimate of the filtered target 
\{$\hat y_i$\}
\begin{eqnarray}
\hat y_i = \eta_i - a\sin\Phi_i \hskip 2mm ; \hskip 2mm  
a={2\over n}\sum_{j=1}^n \eta_j\sin\Phi_j \hskip 2mm .
\end{eqnarray}
It is seen that the amplitude of the induced periodic signal is expected 
to be very small, because it is computed through the Fourier 
transformation of the noise. 

In order to compute the power spectrum of \{$\hat y_i$\} at arbitrary 
test frequency $\omega$, we take the Fourier transform of \{$\hat y_i$\}. 
For the sine and cosine transforms we get
\begin{eqnarray}
S = {2\over n}\sum_{i=1}^n \eta_ig_i \hskip 2mm ; \hskip 2mm
C = {2\over n}\sum_{i=1}^n \eta_if_i \hskip 2mm , 
\end{eqnarray}
where
\begin{eqnarray}
f_i = \cos\omega t_i - \alpha \sin\Phi_i \hskip 2mm ; \hskip 2mm 
\alpha = {2\over n}\sum_{j=1}^n \sin \Phi_j \cos\omega t_j \nonumber \\
g_i = \sin\omega t_i - \beta \sin\Phi_i \hskip 2mm ; \hskip 2mm 
\beta = {2\over n}\sum_{j=1}^n \sin \Phi_j \sin\omega t_j \hskip 2mm .
\end{eqnarray}
By using the properties of equation (A1), we get for the expectation 
value of the power $P(\omega)=S^2+C^2$ 
\begin{eqnarray}
E(P(\omega)) = {4\sigma^2\over n^2}
\Bigl[n-(\alpha^2+\beta^2)(n-\sum_{i=1}^n \sin^2\Phi_i)\Bigr] \hskip 2mm .
\end{eqnarray}
By omitting sums of oscillating terms, far from the template frequency 
$\omega_0$ we get the well-known result for the average noise power (see, 
e.g., Kov\'acs 1980)
\begin{eqnarray}
E(P(\omega)) = {4\sigma^2\over n} \hskip 2mm . 
\end{eqnarray}
At the template frequency, with the above assumption on the oscillating 
terms, we have $\alpha^2+\beta^2\approx 1$ and 
$\sum_{i=1}^n \sin^2\Phi_i \approx n/2$. Therefore, we get  
\begin{eqnarray}
E(P(\omega_0)) = {2\sigma^2\over n} \hskip 2mm . 
\end{eqnarray}
Thus, $E(P(\omega_0))/E(P(\omega)) = 1/2$, which 
means that TFA (on the average) {\em does not} induce extra signal 
component in the filtered time series; on the contrary, it `whitens out' 
(in the statistical sense) even the pure noise target, resulting 
in a decrease of a factor of two in the average power at the 
template frequency.


\label{lastpage}

\end{document}